\documentstyle[12pt]{article}
\makeatletter
\def\kmu{K_L\to\mu\bar{\mu}}

\def\kmn{K^+\to\pi^0\mu^+\nu}

\def\kmu2g{K^+\to\mu^+\nu\gamma}
\newcommand{\nn}{\nonumber}
\newcommand{\bd}{\begin{document}}
\newcommand{\ed}{\end{document}}
\newcommand{\bc}{\begin{center}}
\newcommand{\ec}{\end{center}}
\newcommand{\be}{\begin{eqnarray}}
\newcommand{\ee}{\end{eqnarray}}
\newcommand{\eqn}{\global\def\theequation}
\newcommand{\sw}{sin^2 \theta_W}
\newcommand{\fbd}{f_B}
\renewcommand{\thefootnote}{\alph{footnote}}
\newcommand{\tr}{P_{\perp}}
\newcommand{\kp}{K^+}
\newcommand{\km}{K^-}
\newcommand{\kz}{K^0}
\newcommand{\kzb}{{\bar K}^0}
\newcommand{\pz}{\pi^0}
\newcommand{\pp}{\pi^+}
\newcommand{\pn}{\pi^-}
\newcommand{\mup}{\mu^+}
\newcommand{\mun}{\mu^-}
\newcommand{\kt}{K_{\mu3}}
\newcommand{\kto}{K_{\mu3}^0}
\newcommand{\ktp}{K_{\mu3}^+}
\newcommand{\cph}{C_\gamma}
\newcommand{\cz}{C_Z}
\newcommand{\cb}{C_B}
\newcommand{\zmn}{$\kz \to \pn \mun \nu$}
\newcommand{\zbmn}{$\kzb \to \pp \mup {\bar \nu}$}
\newcommand{\pll}{$\kp \to \pp l {\bar l}$}
\newcommand{\pmn}{$\kp \to \pz \mup\nu$}
\newcommand{\fs}{F_S}
\newcommand{\fp}{F_P}
\newcommand{\fv}{F_V}
\newcommand{\fve}{F_V^{exp}}
\newcommand{\fa}{F_A}
\newcommand{\ub}{\bar \nu}
\newcommand{\mm}{m_{\mu}}
\newcommand{\mk}{m_K}
\newcommand{\mpi}{m_{\pi}}
\newcommand{\sm}{s_{\mu}}
\newcommand{\x}{x_i}
\newcommand{\cpv}{\rlap\slash{CP}}
\newcommand{\lss}{\hat{\plp}.\vec{s}\times \vec{\bar{s}}}
\newcommand{\lbss}{\hat{\pln}.\vec{s}\times \vec{\bar{s}}}
\def\il#1{I_{L}^{#1}}
\def\ilp#1{I_{L}^{\gamma #1}}
\def\ilz#1{I_{L}^{Z #1}}
\def\ilb#1{I_{L}^{B #1}}
\def\li#1{L_{I}^{#1}}
\def\lip#1{L_{I}^{\gamma #1}}
\def\liz#1{L_{I}^{Z #1}}
\def\lib#1{L_{I}^{B #1}}
\def\vv#1{V_{#1 s} V^{*}_{#1 d}}
\def\eps#1#2#3#4{{\varepsilon}^{\mu \nu \rho \sigma} {#1}_{\mu} {#2}_{\nu}
{#3}_{\rho} {#4}_{\sigma}}
\def\sv#1#2{#1\cdot #2}
\def\ep{${e^+}$}
\def\en{${e^-}$}
\def\pk{p_{k}}
\def\ppi{p_{\pi}}
\def\plm{p_{\mu}}
\def\pln{p_{\nu}}
\def\tw#1{{#1}^{+} {#1}^{-}}
\def\twb#1{#1 {\bar #1}}
\def\emu{e^{\mp} {\mu}^\pm}

\def\figcap{\section*{Figure Captions\markboth
     {FIGURECAPTIONS}{FIGURECAPTIONS}}\list
     {Figure \arabic{enumi}:\hfill}{\settowidth\labelwidth{Figure 999:}
     \leftmargin\labelwidth
     \advance\leftmargin\labelsep\usecounter{enumi}}}
\let\endfigcap\endlist \relax
\def\reflist{\section*{References\markboth
     {REFLIST}{REFLIST}}\list
     {[\arabic{enumi}]\hfill}{\settowidth\labelwidth{[999]}
     \leftmargin\labelwidth
     \advance\leftmargin\labelsep\usecounter{enumi}}}
\let\endreflist\endlist \relax

\renewcommand{\thefootnote}{\alph{footnote}}
\begin{document}
\tolerance=10000
\begin{titlepage}
\begin{flushright}
{\normalsize     NHCU-HEP-94-03\\
hep-ph/9410347
}
\end{flushright}
 \null
 \vskip 0.5in
\begin{center}
 \vspace{.15in}
{\Large {\bf Analysis of Transverse Muon Polarization in }}\\
{\Large {\bf  $\kmn$ and $\kmu2g$ }}\\
{\Large {\bf Decays with Tensor Interactions  }}
  \par
 \vskip 2.5em
 {\large
  \begin{tabular}[t]{c}
        {\bf C.~Q.~Geng$^a$ and S.~K.~Lee$^b$}
\\
\\
   {\em   $^a$Department of Physics, National Tsing Hua University} \\
  {\em  Hsinchu, Taiwan, Republic of China} \\
and\\
{\em $^b$Department of Physics, Iowa State University} \\
{\em Ames, IA 50011 USA}
   \end{tabular}}
 \par \vskip 5.0em
 {\large\bf Abstract}
\end{center}
\setlength{\baselineskip}{5ex}

The T violating transverse muon polarizations in
$\kmn$ and $\kmu2g$ decays due to tensor interactions are studied.
The magnitudes of these polarizations over the allowed phase
space are presented.

\vskip 5cm

\end{titlepage}

\setlength{\baselineskip}{5ex}
\pagestyle{plain}
\pagenumbering{arabic}
\setcounter{page}{2}

Although the standard model has been an enormous success in explaining
experimental data, it is generally anticipated that there is
new physics in higher energy regions. One of the long-shot efforts
to exploring such new physics could be searches for CP violation
(or T violation) outside of
the Cabibbo-Kobayashi-Maskawa (CKM) paradigm \cite{ckm}.

It is well known that measuring a component of muon polarization normal
to the decay plane in $\kmn$ \cite{Sakurai} or $\kmu2g$ \cite{Bryman}
decays would signal T violation.
These muon polarizations, called transverse
muon polarizations ($P_{\perp}$), are related to the T-odd triple correlations:
\be
\vec{s}_{\mu}\cdot (\vec{p}_{\mu}\times \vec{p}_{\pi})
\ \ \ \ \ {\rm and} \ \ \ \ \
\vec{s}_{\mu}\cdot (\vec{p}_{\mu}\times \vec{p}_{\gamma})
\ee
for the decays $\kmn$ and $\kmu2g$, where $\vec{s}_{\mu}$ is
the muon spin vector and $\vec{p}_{i}\ (i=\mu\,,\ \pi$ and $\gamma$)
represent the momenta of the muon, pion and photon
in the rest frame of $K^+$, respectively.
It has been shown \cite{Leurer,bg} that, for $\kmn$ decay,
the transverse muon polarization defined by Eq. (1) is equal to zero in
any theory of CP violation
with the decay process via intermediate vector bosons
(including the standard CKM model).
It is also expected \cite{geng-kek} that the CKM phase does not induce
the muon polarization in $\kmu2g$ decay.
Therefore measurements of these polarizations
could be clear signatures of physics beyond the standard model.
There are a number of different sources that might give rise to these
polarizations, the most important ones being the weak CP violation
from some kinds of non-standard CP violation models.
The electromagnetic interaction among the final state particles
can also make contributions, which are usually less interesting
and could even hide the signals from the weak CP violation \cite{FSI}.
We shall refer to these final-state-interaction (FSI) contributions
as theoretical backgrounds.

In this paper we will study the transverse muon
polarizations in $\kmn$ and $\kmu2g$ decays in the precence of
tensor interactions. We emphasize that these tensor interactions
would undoubtedly be signals of new physics.
Recently there has been considerable interest in the possibility of having
tensor interactions in weak decays in connection with the experiments on
$\pi\to e^-\bar{\nu}\gamma$ \cite{pi-T} and $K^+\to\pi^0 e^+\nu$
\cite{ke3-T}, in which tensor interactions have been introduced
to explain the data. An analogous tensor interaction in
$K^+\to l^+\nu\gamma\ (l=e\,,\ \mu)$
decays was discussed in Ref. \cite{Gabrielli}.
For the decay $\kmn$, we remark that the muon polarization induced
by the tensor intreractions have been investigated previously,
for example, in Refs. \cite{Leurer,bg}. Here we shall give a detail
analysis on distributions of the transverse muon polarizations
in terms of Dalitz plots and derive general constraints on the form factors.
For the completeness, we will also include the scalar interactions in our
discussions on $P_{\perp}(\kmn)$.

We start by writing the decays as
\be
K^+ (p_K) \to \pi^0({p}_{\pi})\;\mu^+({p}_{\mu},{s}_{\mu})\;\nu
({p}_{\nu})\;,
\ee
and
\be
K^+ (p_K)  \to  \mu^+({p}_{\mu},{s}_{\mu})\;\nu({p}_{\nu})\;\gamma(p_{\gamma})
\ee
where ${p}_{i}\ (i=K\,,\ \pi\,,\ \mu \,,\ \nu\,,\ \gamma)$
 are the four-momenta and ${s}_{\mu }$
is the four-spin muon polarization vector, respectively.
For the decay $\kmn$,
we use the most general invariant amplitude adopted by the
Particle Data Group \cite{pdg}
\be
{\cal M} &=&{\cal M}^{SM}+{\cal M}^{NSM}
\ee
where
\be
{\cal M}^{SM}&=&{G_F\over 2} sin\theta_c[f_+(q^2)(\pk+\ppi)^{\alpha}+
f_-(q^2)(\pk-\ppi)^{\alpha}]\bar{\mu}\gamma_{\alpha}(1-\gamma_5)\nu
\ee
and
\be
{\cal M}^{NSM} &=&G_Fsin\theta_c[M_Kf_S\bar{\mu}(1-\gamma_5)\nu+
{f_T\over M_K}p^{\alpha}_Kp^{\beta}_{\pi}
\bar{\mu}\sigma_{\alpha\beta}(1-\gamma_5)\nu]
\ee
are the amplitudes from the standard and
non-standard interactions.
Here $f_S$ and $f_T$, known as the scalar and tensor form factors,
are related to the effective scalar and tensor interactions, respectively
and we have included the non-standard scalar part in the amplitude
for the completeness.
We note that in the case of $K^+\to \pi^0e^+\nu$ decay, a fit of the
experiment \cite{ke3-T}
gives that $|f_S/f_+|=0.070\pm0.016$ and
$|f_T/f_+|=0.53\begin{array}{c} +0.09\\ -0.10 \end{array}$.
These experimental values could also suggest large interactions
of Eq. (6) in $\kmn$ decay, which indeed have not been ruled out from the
experimental data \cite{pdg}.

The probability of the decay as a function of the 4-momenta of the
particles and the polarization 4-vector $\sm$ of the muon can be
written as
\be
dw=(1+\sv{\sm}{P})\Phi'\rho/(2E_K)
\ee
where $\Phi'$ is a phase space factor,
\be
\rho={d\vec{p}_{\pi}d\vec{p}_{\mu}d\vec{p}_{\nu}\over 2E_{\pi}2E_{\mu}2E_{\nu}}
\delta^4(p_K-\ppi-\plm-\pln) (2\pi)^{-5}\,,
\ee
and P is the 4-vector muon polarization.
In the kaon rest frame, the transverse component of P, i.e.,
transverse muon polarization, is given by
\be
\vec{P}_{\perp}&=&
\left(2Imf_S+{1\over M_K^2}[\mm^2-2M_K(E_{\mu}-E_{\nu})]Imf_T\right)
{\vec{p}_{\pi}\times\vec{p}_{\mu}\over \Phi}
\ee
with
\be
\Phi &=&(M_K-E_{\pi}-E_{\mu})\left(2E_{\mu}-{\mm^2\over M_K}\right)
\nn\\
&&-{1\over 2}(M_K^2+M_{\pi}^2-\mm^2-2E_{\pi}M_K)
\left(1-{\mm^2\over 4M_K}\right)\,,
\ee
where we have used $f_+\simeq 1$ and $f_-\simeq 0$.
Clearly,
the non-zero contributions to
$P_{\perp} \equiv |\vec{P}_{\perp}| $ in $\kmn$ could arise if
there are some non-standard effective scalar and/or tensor interactions.

At the present the best bound on the transverse muon polarization
in Eq. (9) comes from the
experiment at BNL  \cite{BNL} with the value of
$(-3.1\pm5.3)\times 10^{-3}$,
leading to the upper limit being about $10^{-2}$ at $90\%$ C.L..
In the absence of the tensor form factor in Eq. (9),
contours for the magnitude of the transverse
muon polarization, $P_{\perp}$, are shown in Figure 1.
Similar contours with assuming $f_S=0$ are depicted in Figure 2.
  From $P_{\perp}^{expt}(\kmn) <10^{-2}$ and
Figures 1 and 2,
we find that
\be
|Imf_S|\leq 10^{-3}\ \ \ \ \ {\rm and}
\ \ \ \ \
|Imf_T|\leq 10^{-3}\,.
\ee
Therefore, the transverse muon polarization in $\kmn$ could
be  $10^{-2}$ without conflicting with the
experimental constraints.

The muon polarization effects from these effective non-standard structures
could be accessible to the underway experiment of E246
at KEK \cite{E246,kuno}, in which a sensitivity
of $0.05\%$ in $P_{\perp}(\kmn)$ will be obtained,
and future experiments in a kaon factory \cite{kuno_k} as well.
The well known examples which could
give rise to the effective scalar interactions
in Eq. (6) are the multi-Higgs and leptoquark
models \cite{bg,Cheng,Kane,Leurer},
while the tensor ones could be induced from leptoquark models \cite{bg,Leurer}.

Finally we remark that
the theoretical backgrounds, i.e., {\em FSI} contributions,
to the transverse muon polarization,
can be ignored
since they are expected to be $O(10^{-6})$
arising from two-loop diagrams \cite{zhitnitskii}.

We now study the decay of $\kmu2g$.
In the framework of the standard model, the amplitude of the decay
can be written as
\be
{\cal M}& =& {\cal M}_{IB}+{\cal M}_{SD}\,
\ee
where ${\cal M}_{IB}$ and ${\cal M}_{SD}$ are the so called
{\em inner bremsstrahlung} (IB) and {\em structure dependent}
(SD) terms, given by
\be
{\cal M}_{IB}={ieG_F\over \sqrt{2}}\sin\theta_cm_{\mu}
f_K\epsilon_{\alpha}^*\bar{u}(p_{\nu})(1-\gamma_5)
\left({p^{\alpha}\over pq}-
{q\gamma^{\alpha}+2p_{\mu}^{\alpha}\over 2p_{\mu}q}\right)v(p_{\mu})
\ee
and
\be
{\cal M}_{SD} &=&
{ieG_F\over \sqrt{2}}\sin\theta_c\epsilon_{\alpha}^*\left[
pq{F_A\over M_K}({p^{\alpha}q^{\beta}\over pq}-g^{\alpha\beta})
-i\epsilon_{\alpha\beta\rho\lambda}
{F_V\over M_K}p^{\rho}q^{\lambda}\right] \nn\\
&& \bar{u}(p_{\nu})\gamma_{\beta}(1-\gamma_5)v(p_{\mu})\,.
\ee
Here $\epsilon_{\alpha}$ is the photon polarization vector, $f_K$ is the K
decay
constant, and $F_{A,V}$ are the axial, vector form factors defined by
\be
<\gamma|\bar{u}\gamma_{\alpha}\gamma_5s|K> &=&
-ie{F_A\over M_K}\epsilon_{\beta}(pqg_{\alpha\beta}-p_{\beta}q_{\alpha})
\nn\\
<\gamma|\bar{u}\gamma_{\alpha}s|K> &=&
e{F_V\over M_K}\epsilon_{\alpha\beta\rho\lambda}
\epsilon^{\beta}q^{\rho}p^{\lambda}\,,
\ee
respectively.
At the one-loop level in chiral perturbation theory, the form factors
$F_{A,V}$ are found to be \cite{Ecker}
\be
&&F_V=-0.0945\,,\ \
F_A=-0.0425\,
\ee
which agree with experiments.
It is easily seen that the standard matrix elements
in Eqs. (13) and (14) do not generate the T-odd triple correlation
term in Eq. (1) for $\kmu2g$ decay
because there is no relative phase between them at the tree level.
It is clear that to have non-zero transverse muon polarization, a
non-standard matrix element of new physics is needed. One possible
candidate for such new physics in $\kmu2g$ is to have
a tensor interaction given by
\be
{\cal L}^{T}_{eff}=
{G_F\over \sqrt{2}}\sin\theta_cf_T'\bar{u}\sigma_{\alpha\beta}s
\bar{\mu}\sigma^{\alpha\beta}(1-\gamma_5)\nu
\ee
where $f_T'$ is a constant form factor.
The possibility of having the interaction in Eq. (17) for $\kmu2g$
has been  explored recently in Ref. \cite{Gabrielli}
motivated by the experimental results \cite{pi-T}.
With some arbitrary hypotheses \cite{Gabrielli}
and results from the pion decays with PCAC approximation
\cite{Kogan},
the form factor $f_T'$ is expected
to be less than $6.2\times 10^{-2}$.
 From Eq. (17), one obtains \cite{Gabrielli}
\be
{\cal M}^{T}&=&
{ieG_F\over \sqrt{2}}\sin\theta_c{\epsilon^{\alpha}}^*q^{\beta}F_T
\bar{u}(p_{\nu})\sigma_{\alpha\beta}(1+\gamma_5)v(p_{\mu})
\ee
where the form factor $F_T$ is defined by
\be
<\gamma|\bar{u}\sigma_{\alpha\beta}\gamma_5s|K>=
-ie{F_T\over f_T'}{1\over 2}
(\epsilon_{\alpha}q_{\beta}-\epsilon_{\beta}q_{\alpha})
\ee
with the form factor $F_T$ being
\be
|F_T|<2.2\times 10^{-2}\,
\ee
assumed in \cite{Gabrielli}.
We note that the tensor interaction in Eq. (17) in principle could
also lead to a contribution to $\kmn$ decay \cite{Leurer}.
However, the matrix element corresponding to Eq. (17)
is suppressed by PCAC \cite{Gabrielli} unlike the relation
in Eq. (19) for the case of $\kmu2g$ mode.
We therefore believe that the decay of $\kmu2g$ is a more interesting
one than $\kmn$ to probe the tensor interaction of Eq. (17).

Similar to the discussions of
$P_{\perp}(\kmn)$, the transverse muon polarization
for $\kmu2g$ arising from the interferences between
the tensor term in Eq. (18) and the IB and SD terms in Eqs. (12)-(14)
is found to have the following form
\be
\vec{P}_{\perp}&=&{4(1-\lambda)\over x}
\left[(F_V-F_A)x^2-{2\mm^2f_K\over M_K^3\lambda}\right]
\cdot ImF_T{\vec{p}_{\mu}\times \vec{p}_{\gamma}\over M_K^2}\cdot
{1\over \rho (x,y)}
\ee
with
\be
x& =& {2E_{\gamma}\over M_K}\,,\ \
y\:=\:{2E_{\mu}\over M_K}\,, \nn\\
\lambda&=&(x+y-1-r_{\mu})/x\,,\ r_{\mu}\:=\:{\mm^2\over \mk^2}\,,
\ee
and $\rho (x,y)$ is the normalized Dalitz plot density
given by
\be
\rho(x,y)=
\rho_{\mbox{\tiny{IB}}}(x,y) +  \rho_{\mbox{\tiny{SD}}}(x,y)
+  \rho_{\mbox{\tiny{IBSD}}}(x,y)\nn\\
+\rho_{\mbox{\tiny{T}}}(x,y)  +  \rho_{\mbox{\tiny{IBT}}}(x,y)
+  \rho_{\mbox{\tiny{SDT}}}(x,y)
\ee
where
\be
 \rho_{\mbox{\tiny{IB}}}(x,y) &=&
2r_{\mu}\left({f_K\over M_K}\right)^2f_{\mbox{\tiny{IB}}}(x,y)
\nonumber \\
 \rho_{\mbox{\tiny{SD}}}(x,y)&=&
{1\over 2}\left[ (F_V+F_A)^2
f_{{\mbox{\tiny{SD}}^+}}(x,y)
+(F_V-F_A)^2f_{{\mbox{\tiny{SD}}^-}}(x,y)\right]
\nonumber \\
 \rho_{\mbox{\tiny{IBSD}}}(x,y) &=&
2r_{\mu}{f_K\over M_K}
\left [ (F_V+F_A) f_+(x,y)+ (F_V-F_A)f_-(x,y)\right]
\nonumber \\
\rho_{\mbox{\tiny{T}}}(x,y) &=& 2|F_T|^2f_{\mbox{\tiny{TT}}}(x,y)
\nonumber \\
\rho_{\mbox{\tiny{IBT}}}(x,y)
&=& 4\sqrt{r_{\mu}}{f_K\over M_K}Re(F_T)f_{\mbox{\tiny{IBT}}}(x,y)
\nonumber \\
\rho_{\mbox{\tiny{SDT}}} &=& 2\sqrt{r_{\mu}}Re(F_T)(F_V-F_A)
f_{\mbox{\tiny{SDT}}}(x,y)
\ee
with
\be
f_{\mbox{\tiny{IB}}}(x,y) &=&
\frac{1-y+r_{\mu}}{\lambda x^3}
\left(x^2 +2(1-x)(1-r_{\mu})
-\frac{2r_{\mu} (1-r_{\mu})}{\lambda} \right)
\nn\\
 f_{{\mbox{\tiny{SD}}^+}}(x,y) &=& x\lambda [(\lambda x+r_{\mu})(1-x)
-r_{\mu}]
\nn\\
 f_{{\mbox{\tiny{SD}}^-}}(x,y)&=& x(1-\lambda) [(1-x) (1-y) +r_{\mu}]
\nn\\
 f_+(x,y) &=& {1-\lambda\over x\lambda}
\left[ (1-x)(1-x-y)+r_{\mu} \right]
\nn\\
 f_-(x,y)& =&{(1-\lambda)x\over \lambda}-f_+(x,y)
\nn\\
f_{\mbox{\tiny{TT}}}(x,y)&=&\lambda x^2 (1-\lambda)
\nn\\
f_{\mbox{\tiny{IBT}}}(x,y)&=&1+r_{\mu}-\lambda-{r_{\mu}\over \lambda}
\nn\\
f_{\mbox{\tiny{SDT}}}(x,y)&=&\lambda x^2 (1-\lambda)\,.
\ee
In Figure 3, we show the contours for the
magnitude of the transverse muon polarization
in Eq. (21) induced by the tensor interaction.
 From the constraint in Eq. (20) we conclude that
$P_{\perp}(\kmu2g)$ can be as large as $10\%$.

Although there is only one charged final state particle for
the decay $\kmu2g$ like $K_{\mu 3}^+$ mode, the FSI due to
electromagnatic interaction arises at one-loop diagrams because of the
existence of the photon in the final state.
Therefore it is expected \cite{Marciano}
that the theoretical background, i.e., FSI,
for the polarization in Eq. (21) is large,
unlike the case in $K_{\mu 3}^+$ where the FSI is at the two-loop level.
Recently, we have performed \cite{gl}
a detail calculation on the FSI contribution to the polarization
and found that $P^{FSI}_{\perp}(\kmu2g)$ is at the level of
$10^{-3}$ in most regions of the decay allowed phase space.

Finally we note that the experiment E246 at KEK could also measure
the muon polarization \cite{kuno} in Eq. (21).
Sensitivity at the level of $10^{-3}$ may be possible \cite{kuno_p}.
This will be a useful calibration for the experiment.

In summary, we have examined the T violating transverse muon polarizations
in $\kmn$ and $\kmu2g$ decays in the presence of non-standard interactions
such as the tensor interactions.
We have shown that the polarizations
are expected to be large without conflicting with the current
experimental data
and they could be accessible at future experiments such as the underway
experiment of E246 at KEK.
Measuring the transverse muon polarizations in $\kmn$ and $\kmu2g$
will be a clear indication of physics beyond the standard model
and provide some insights into the origin of CP violation.
In particular, these measurements could indicate
the existence of the tensor interactions.

\newpage
\noindent
{\bf Acknowledgments}

The author would like to thank P. Depommier, J. Imazato,
T. Inagaki, Y. Kuno, W. Marciano, and S. Sugimoto
for discussions. This work was supported in part
by the National Science Council of Republic of China
under Grant No. NSC-83-0208-M-007-118.

\newpage

\begin{figcap}

\item
Dalitz plot of $P_{\perp}(\kmn)$ for $Imf_T=0$ with
the values of contours: (a) - (e) being 0,
0.8, 1.4, 2.4 and 6.0 in the unit of $|Imf_S|$, respectively.
\item
Dalitz plot of $P_{\perp}(\kmn)$ for $Imf_S=0$ with
the values of contours: (a) - (f) being 0.2
0.1, 0, 0.3, 1.0, and 4.0 in the unit of $|Imf_T|$, respectively.
\item
Dalitz plot of $P_{\perp}(\kmu2g)$.
The contours (a) - (d) are 0.5, 1.5 3.0 and 6.0 in the unit of
$|ImF_T|$, respectively.

\end{figcap}

\ed